\documentclass{article} 
\usepackage{iclr2021_conference,times}

\usepackage{amsmath,amsfonts,bm}

\def\eqref#1{equation~\ref{#1}}

\def\1{\bm{1}}

\DeclareMathAlphabet{\mathsfit}{\encodingdefault}{\sfdefault}{m}{sl}
\SetMathAlphabet{\mathsfit}{bold}{\encodingdefault}{\sfdefault}{bx}{n}

\usepackage{hyperref}
\usepackage{url}
\usepackage{graphicx}

\title{Efficient data selection methods for the \\ development of machine learned potentials}

\author{Jan Finkbeiner, Samuel Tovey \& Christian Holm  \\
Institute for Computational Physics\\
University of Stuttgart\\
Allmandring 3, 70569, Stuttgart, DE \\
{\{jfinkbeiner, stovey, holm\}@icp.uni-stuttgart.de} \\
}

\iclrfinalcopy 
\begin{document}

\maketitle

\begin{abstract}
We present an investigation into data selection methods for the efficient sampling of configuration space as applied to the development of inter-atomic potentials for scale bridging in molecular dynamics (MD) simulations. This investigation suggests that the most efficient sampling techniques are those that incorporate information on an atomic level such as forces or atomic energies. Finally, we generate an inter-atomic potential for the a sodium chloride system using each data selection technique and find that the global selection methods result in non-physical simulations.
\end{abstract}

\section{Introduction}
\label{sec:introduction}
In the development of any supervised regression model it is important to consider how to generate training data to optimally represent your target function. In the case where generation of this data is expensive, it is further necessary to considering how to minimize the number of points required for effective training. One application where this is especially prevalent is the generation of inter-atomic potentials for use in molecular dynamics (MD) simulations. In these applications, expensive ab-initio data is generated on small atomic system over short time scales before a machine learning algorithm such as Gaussian process regression (GPR) (\cite{bartok10a}) or a neural network (NN) (\cite{behler07a, schuett18a}) is used to fit a function that maps atomic environments to a system energy and atomic forces.
Once constructed, these potentials can be used for classical MD simulations and employed on larger atomistic systems for longer time scales. In these simulations, the machine learned models can achieve near ab-initio accuracy with a computational complexity of $\mathcal{O}(N)$ where N is the number of atoms in the system (\cite{tovey20a, sivaraman20a}). The fitting procedure involves the deconstruction of the global energy into contributions from atomic environments. In order to do this, the atomic environments are transformed into so-called descriptors that are then passed into a machine learning algorithm. These descriptors encode symmetries present in the potential such as rotation, translation and particle exchange. In order to develop an effective model, it is necessary that the training data contain as many unique atomic environments as possible such that one maximally samples configuration space. Whilst this may be stated simply, it is not clear in practice what constitutes a unique atomic environment, or how best to assess an environment for uniqueness.
 There are a number of factors in the development of these inter-atomic potentials which are currently being investigated including the choice of descriptor (\cite{bartok13a, behler07a, behler11a, gastegger18a, lindsey17a, rupp12a, samanta18a, seko14a, shapeev16a, takahashi18a, thompson15a, zhu16a}), or the machine learning algorithm used in the fitting procedure (\cite{rupp12a, schuett17a, balabin11a, bartok10a, behler07a}). In the case of data selection for these models, it is often the case that training data is sampled uniformly in time from long, expensive, ab-initio MD simulations (\cite{cole20a,shao20a}), with some notable steps being taken in the direction of active learning (\cite{sivaraman20a}), descriptor space metrics (\cite{de16a}), and in the direct manipulation of atomic structure to induce rare events as in the so-called RAG sampling procedure (\cite{choi20a}).

In this work, several approaches to selecting data for use in the development of an inter-atomic potential are presented and tested in order to understand what factors might impact model performance. The methods studied incorporate both global properties of a configuration, as well as the local, atomic information, in order to understand which approach leads to a more accurate model. Data selected by each method is used to train an inter-atomic potential on simulated molten sodium chloride which is then assessed based on the root mean square error (RMSE), maximum error, and mean absolute error (MAE) of the model predictions. 

\section{Data Selection Methods}
\label{sec:data_selection_methods}

\begin{figure}
\centering
	\includegraphics[width=1.0\textwidth]{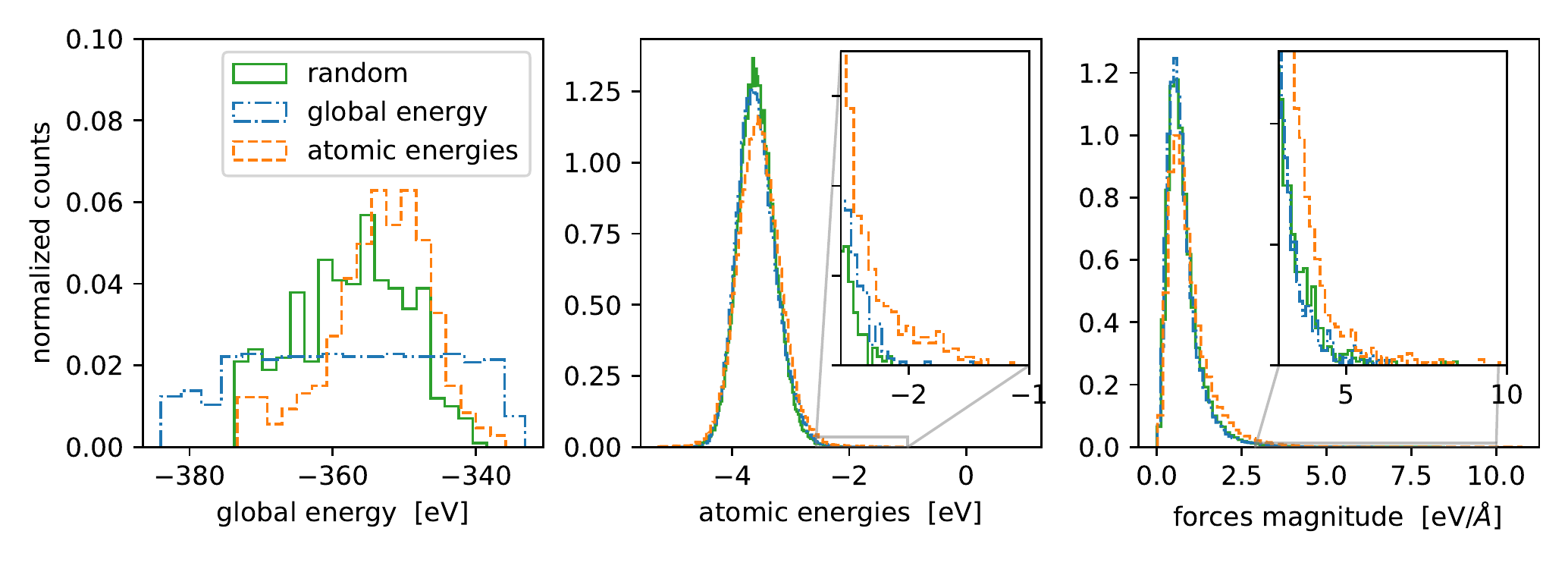}
	\caption{Resulting distributions of global energy (left), atomic energies (middle) and force magnitude (right) in data-sets for different selection methods presented in Section \ref{sec:data_selection_methods}. Each data-set consists of 512 configurations selected via the random-selection method (solid, green), sampling uniformly in global energy (dash-dotted, blue), or sampling uniformly in atomic energies (dashed, orange). Insets show zoom-ins in the long-tail of the distributions.}
	\label{fig:data_distribution}
\end{figure} 
In training an inter-atomic potential on global energy data we make an underlying assumption that this energy may be decomposed into atomic contributions as
\begin{align}
    \centering
    \label{eqn:MLPES_Assumption}
    E = \sum\limits_{i}^{N}\epsilon_{i},
\end{align}
where $\epsilon_{i}$ is the contribution of the $i^{\text{th}}$ atomic environment and N is the number of atoms in the system.  As was briefly mentioned in the introduction, the key to data selection in the development of machine learned inter-atomic potentials is sufficient sampling of configuration space such that all unique configurations resulting in some $\epsilon_{i}$ above are realised in the training data. In order to ensure that large parts of configuration space are sampled, we perform several MD simulations over different sets of constant parameters, in our case, fixed atom number, pressure, and temperature. On this pool of samples we apply one of the selection methods below to identify configurations that contain relevant atomic environments.

\subsection{Random Selection}
In the random sampling approach, configurations over the full MD trajectory are sampled at random and used as training, test, and validation data. As a benchmark for the performance of the sampling methods over different simulations, we also use this method on data from a single MD simulation which we refer to as single-MD random sampling.

\subsection{Global Energy Selection}
In ab-initio simulations, the Schr\"{o}dinger equation is solved numerically using density functional theory (DFT) (\cite{burke12a}) in order to determine the total energy of the system of atoms. It is on this energy, along with the forces on each of the atoms, that the machine learning algorithm employed will train. Therefore, it is rational to select training data by sampling uniformly across the energy values as illustrated in Figure \ref{fig:data_distribution}. Whilst this approach ensures the existence of unique global energies in the training data, it is not necessarily true that these configurations contain within them unique atomic environments.

\subsection{Atomic Energy Selection}
Whilst in an ab-initio simulation the concept of a atomic energy in Equation \ref{eqn:MLPES_Assumption} is controversial, in a classical simulation it can be written simply as the summation of terms in the inter-atomic potential calculation up to a defined cutoff as
\begin{align}
    \centering
    \label{eqn:classical_energy}
    \epsilon_{i} = \sum_{j}^{N_{\text{pairs}}}U(r_{ij}, r_{c}),
\end{align}
where U is a function determining the potential energy between two atoms $i$ and $j$, $r_{ij}$ is the pairwise distance between the atoms $i$ and $j$, $r_{c}$ is the short range cutoff of the potential, and the summation runs over all atoms forming an interacting pair with atom $i$. Due to the similarity between Equation \ref{eqn:classical_energy} and the fundamental assumption made in Equation \ref{eqn:MLPES_Assumption}, this is a candidate for data selection. In our study, we generate data-sets by selecting configurations uniformly based on atomic energies.

\subsection{Force Selection}
Along a similar line of thinking to the previous method, force sampling also looks to an atomic property to choose interesting configurations. In this case, there is the added benefit of forces being quantum mechanical observables, and therefore available in an ab-initio simulation through the Hellmann-Feynman theorem (\cite{guettinger32a, feynman39a}). In this method, rather than the atomic energy being studied, the net force on the $i^{\text{th}}$ atom is used to indicate a unique environment.

\section{Results}

\begin{figure}[ht!]
\centering
\includegraphics[width=1.0\textwidth]{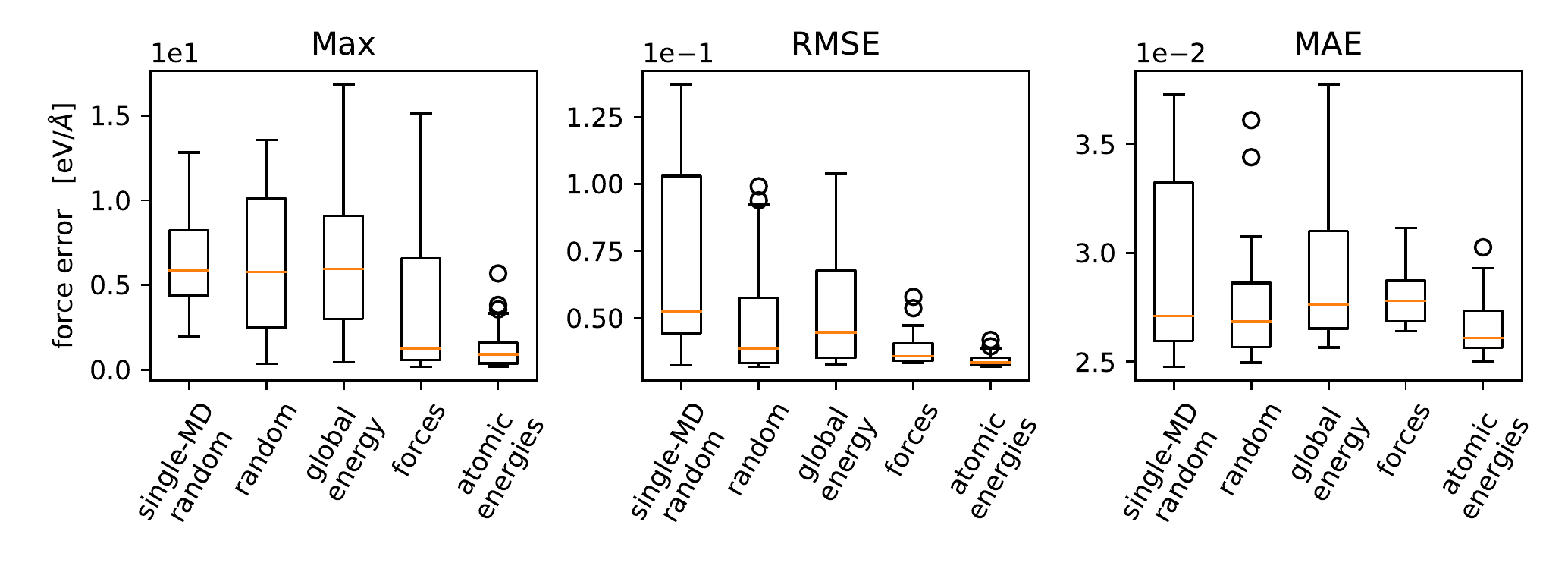}
\caption{Comparison of errors in force predictions for NN-models trained on data-sets that were selected by the different methods presented in Section \ref{sec:data_selection_methods}. From left to right we show plots for the maximum error (Max), root-mean-square error (RMSE), and the mean absolute error (MAE) in the ML models. } 
\label{fig:classical_force_errors}
\end{figure}

In order to test the effectiveness of each selection method data was generated from 100 atom classical MD simulations of molten sodium chloride using a Born-Meyer-Huggins-Tosi-Fumi (BMHTF) potential (\cite{tosi64a, fumi64a, mayer33a, born32a, huggins33a}) for temperatures and pressures ranging from 900 K to 2200 K and $1 \cdot 10^{-3}$ bar to $5\cdot 10^{4}$ bar respectively.

We initially evaluated the performance of the selection methods by training models on data chosen by each of them and then calculating the maximum error (Max), root mean square error (RMSE), and the mean absolute error (MAE) of the model force predictions with respect to test sets. For each selection method, 6 models of differing size and input parameters (see Appendix) were trained. Each model was tested on 20'480 configurations consisting of data-sets spanning different temperatures chosen in equal parts by the selection methods described in Section \ref{sec:data_selection_methods}.

Figure \ref{fig:classical_force_errors} shows a clear trend in the prediction errors. The models trained on data-sets that were generated by the force or atomic energy selection methods achieve lower max and root mean square errors. 
One explanation for the improved performance of the atomic selection approaches might be in their distribution of data. While the global energy selection method appears to produce more diverse global energies, the atomic energy selection method contains these so-called 'fat tails' in the atomic energy distribution. The appearance of these fat tails implies the inclusion of less probable configurations. In the case of global energy selection, the contribution of these atomic energies may be diluted in the summation over all environments, whereas with the atomic approach, they are identified and added to the training data.

\begin{figure}
\centering
	\includegraphics[width=1.0\textwidth]{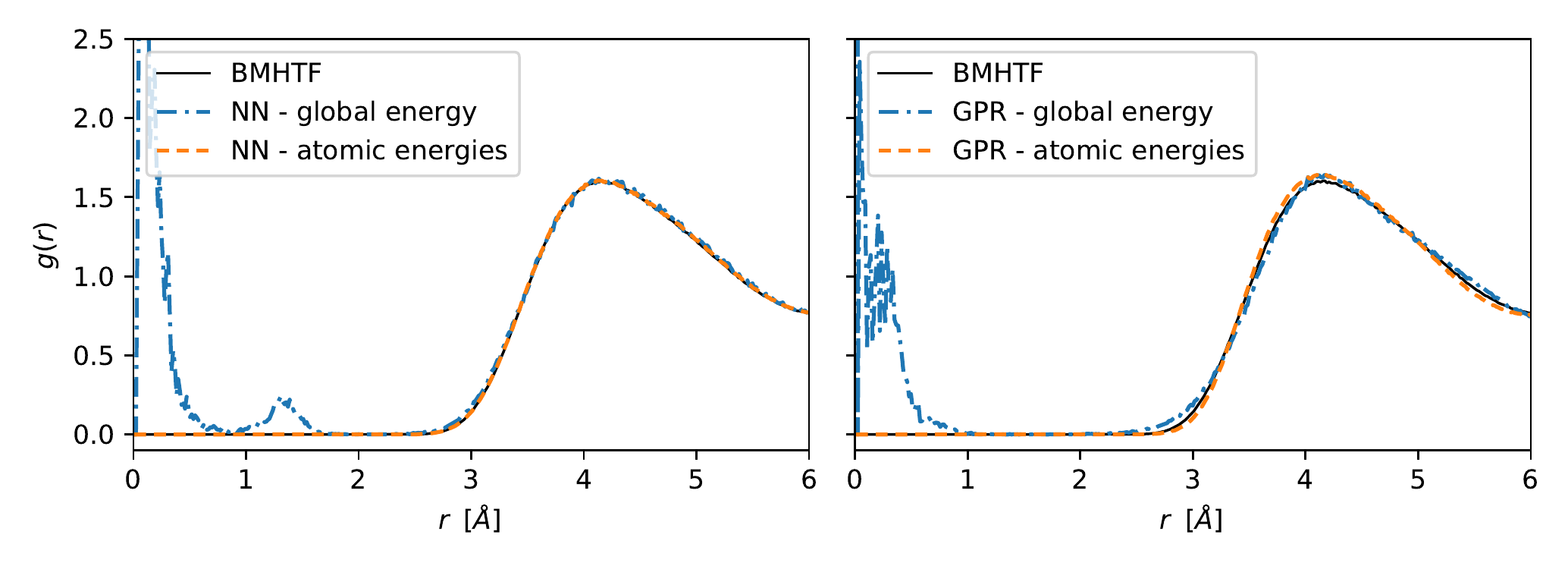}
	\caption{Comparison of the Na-Na radial distribution function computed from a reference BMHTF simulation and NN-driven (left) and GPR-driven (right) MDs with 500 ion-pairs trained on 512 (NN) or 128 (GPR) configurations selected via the global energy and atomic energies selection method.}
	\label{fig:rdfs}
\end{figure} 

As a further validation of the selection techniques, the trained models were used in 1000 atom MD simulations at 1400 K in an NVT ensemble run for up to 1000 ps. This investigation exposes the models to a number of varying configurations from large regions of configuration space thereby assessing robustness. 
In order to eliminate possible algorithm dependence, neural network and Gaussian process regression models were used (see Appendix). The radial distribution functions (RDF) of these simulations were then compared with BMHTF model under the same conditions. We see that, in the case of the global data-selection methods, the radial distribution functions contain non-physical short range peaks implying the training data did not sufficiently represent the potential energy surface. In the case of the atomic energies however, no such short range peaks arise and the function fits that of the reference BMHTF potential. The success of the atomic property selection methods highlights the robustness that accompanies the low RMSE and Max errors seen in Figure \ref{fig:classical_force_errors}.

\section{Conclusion}
We have shown that data selection based on atomic energies or forces yields more accurate inter-atomic potentials than those trained on global energies or randomly chosen configurations. We found that the calculated RMSE values of models trained using the atomic energies data selection method was consistently lower than any other approach. Furthermore, potentials trained on global properties produced non-physical results when used in an MD simulation, whereas those trained on atomic properties reproduced the reference data.
These results suggest that selecting data based on atomic properties results in more accurate and robust machine learned potentials as opposed to global property selection methods.

\subsubsection*{Acknowledgments}
The authors  acknowledge financial support from the German Funding Agency (Deutsche Forschungsgemeinschaft DFG) under Germany’s Excellence Strategy EXC 2075-390740016, and S. Tovey was supported by a LGF stipend of the state of Baden-Württemberg

\newpage
\bibliography{iclr2021_conference}
\bibliographystyle{iclr2021_conference}

\newpage
\appendix
\section{Appendix}

\subsection{SchNet-Hyperparameters}

The Neural Network models are based on the SchNet-architecture (\cite{schuett17b}, \cite{schuett17a}). All hyperparameters used for the different model sizes are summarized in Table \ref{tab:schnet_hyperparams} and \ref{tab:schnet_hyperparams_train}. If not specified, the default value was chosen.

\begin{table}[h!]
\caption{Hyperparameters of the SchNet-Model types}
\label{tab:schnet_hyperparams}
\begin{center}
\begin{tabular}{c|c|c|c}
\multicolumn{1}{c}{\bf NAME}  & \multicolumn{1}{c}{\bf Small} & \multicolumn{1}{c}{\bf Medium} & \multicolumn{1}{c}{\bf Large}
\\ \hline \\
    \texttt{n\_interactions}  &3            &4                  &4\\
    \texttt{n\_atom\_basis}   &32           &48                 &64\\
    \texttt{n\_filters}       &32           &64                 &128\\
    \texttt{n\_gaussians}     &3            &48                 &64\\
    \texttt{n\_in}            &32           &48                 &64\\
    \texttt{elements}         &(11, 17)     &(11, 17)           &(11, 17)\\
    \texttt{n\_neurons}       &[48, 48]     &[128, 64, 32]      &[128, 128, 64, 32]\\
    \texttt{n\_layers}        &3            &4                  &5\\
\end{tabular}
\end{center}
\end{table}

\begin{table}[h!]
\caption{Training hyperparameters and other parameters of all SchNet-Models}
\label{tab:schnet_hyperparams_train}
\begin{center}
\begin{tabular}{c|c}
\multicolumn{1}{c}{\bf NAME}  &\multicolumn{1}{c}{\bf VALUE}
\\ \hline \\
    cutoff radius   &6.0 $\AA$\\
    optimizer       &Adam\\
    learning rate   &5e-4\\
\end{tabular}
\end{center}
\end{table}

\subsection{GAP-Hyperparameters}

To generate the GPR models the GAP suite (\cite{bartok10a}, \cite{bartok13a}) of the QUIP software package was used. GAP suite is available for non-commercial use from www.libatoms.org. The model uses a the SOAP descriptor. Table \ref{tab:SOAP} shows the hyperparameters for the SOAP descriptor.

\begin{table}[h!]
\caption{Hyperparameters of the SOAP descriptor for the GAP model}
\label{tab:schnet_samll_hyperparams}
\begin{center}
\begin{tabular}{c|c}
\multicolumn{1}{c}{\bf NAME}  &\multicolumn{1}{c}{\bf VALUE}
\\ \hline \\
    \texttt{n\_max}                       &8 \\
    \texttt{l\_max}                       &6 \\
    \texttt{atom\_sigma}                  &0.825 \\
    \texttt{zeta}                        &4 \\
    \texttt{cutoff}                      &6.5 \\
    \texttt{cutoff\_transition\_width}     &0.5 \\
    \texttt{delta}                       &1.0 \\
\end{tabular}
\end{center}
\end{table}
\label{tab:SOAP}

\end{document}